\documentclass[12pt]{article}

\usepackage[margin = 1in]{geometry}
\usepackage[round]{natbib}
\usepackage{amsfonts, amsmath, authblk, graphicx, hyperref, xcolor}
\usepackage[title]{appendix}

\DeclareMathOperator*{\var}{var}
\DeclareMathOperator*{\cov}{cov}

\DeclareMathOperator*{\argmin}{arg\,min}

\newtheorem{prop}{Proposition}

\newtheorem{theorem}{Theorem}

\title{Linear shrinkage for predicting responses in large-scale multivariate linear regression}
\author{Yihe Wang}
\author{Sihai Dave Zhao}
\affil{Department of Statistics, University of Illinois at Urbana-Champaign}

\begin{document}

\maketitle
\begin{abstract}
  We propose a new prediction method for multivariate linear regression problems where the number of features is less than the sample size but the number of outcomes is extremely large. Many popular procedures, such as penalized regression procedures, require parameter tuning that is computationally untenable in such large-scale problems. We take a different approach, motivated by ideas from simultaneous estimation problems, that performs linear shrinkage on ordinary least squares parameter estimates. Our approach is extremely computationally efficient and tuning-free. We show that it can asymptotically outperform ordinary least squares without any structural assumptions on the true regression coefficients and illustrate its good performance in simulations and an analysis of single-cell RNA-seq data.
\end{abstract}

\section{Introduction} \label{introduction}
In this paper we study multivariate regression, also known as multi-task learning \citep{zhang2017survey}. We specifically study large-scale problems where the number of outcomes $q$ can be very large. Such problems are common in many different fields, for example in multiple imaging modalities studies \citep{hariri2006imaging}, multiple brain network prediction \citep{calhoun2012multisubject}, multivariate time series prediction\citep{makridakis2018m4}, and modern genomics \citep{gamazon2015gene, gusev2016integrative}. We focus on low-dimensional setting where the number of predictors is smaller than the sample size. This setting frequently arises when predictors are difficult to measure. For example, in Section \ref{real} we consider the problem of imputing the expression of roughly 3,000 genes using a relatively small set of probe genes. Such an imputation procedure can be useful for integrating high-throughput technologies, such as single-cell RNA-seq, with emerging spatial transcriptomic technologies are still relatively low-throughput, such as MERFISH \citep{chen2015spatially}.

Our goal is to predict, for a newly observed $p \times 1$ covariate vector $X_0$, the corresponding unobserved $q \times 1$ outcome vector $Y_0$. We assume our training data consists of $n$ observations following a linear model, where $X$ is an $n \times p$ design matrix with rows $X_i$, $Y$ is an $n \times q$ outcome matrix whose $ik$th entry $Y_{ik}$ is $k$th outcome of the $i$th observation, and
\begin{equation}
 \label{setup}
  Y_{i\cdot}
  =
  X_i^\top B + \epsilon_i,
  \quad
  \epsilon_i
  \sim 
  N(0, \Sigma),
\end{equation}
where $Y_{i\cdot} = (Y_{i1}, \ldots, Y_{iq})^\top$ and $B$ is a $p \times q$ matrix of unknown regression coefficients. We consider $q \gg n$ but assume that $p < n$. We also assume that the design matrix $X$ is the same for all outcomes $k = 1, \ldots, q$, but extension to setups with different design matrices is straightforward.

The literature on multivariate regression is extensive, but a major lesson is that borrowing information across the outcomes in multivariate regression can reduce overall estimation and prediction error. For example, when the outcomes are correlated, seemingly unrelated regression \citep{zellner1962efficient, fiebig2001seemingly} estimates $B$ using weighted ordinary least squares with weights equal to the inverse correlations. The Curds and Whey procedure of \citet{breiman1997predicting} predicts outcomes using a linear combination of other outcomes, and partial least squares regression \citep{wold1975soft, wold2001pls} and general envelope methods \citep{cook2010envelope} perform dimension reduction of both the outcomes and the predictors. However, many existing methods are either unable to improve upon naive methods when the outcomes are independent, or are difficult to implement when the number of outcomes is large.

Penalized regression is arguably the current most popular approach to predicting multivariate outcomes. For example, multiresponse lasso \citep{friedman2010regularization} uses a group lasso penalty that assumes that the outcomes are explained by a common set of a few predictors, and reduced rank regression \citep{velu2013multivariate} uses penalties that assume that $B$ is low rank. However, these methods require tuning parameters that are typically chosen using cross-validation, which is extremely computationally expensive when the number of outcomes is large. Furthermore, for certain penalties like the group lasso, these computations cannot be parallelized across the outcomes.

In this paper we propose a new method that can address this issue. We view the multivariate prediction problem as a simultaneous estimation problem, similar to the classic problem of estimating a vector of normal means \citep{johnstone2019gaussian}. We therefore propose a shrinkage estimator, where we linearly shrink each row of the maximum likelihood estimate of the parameter matrix $B$. We allow the shrinkage factors to depend on the newly observed $X_0$ and learn their optimal values from the data itself, so that no tuning parameters are required. We previously studied a version of this idea where we used nonparametric empirical Bayes methods to learn the optimal form of shrinkage, instead of restricting to linear shrinkage \citep{wang2021nonparametric}. However, because it is nonparametric, this method is computationally challenging, suffers from the curse of dimensionality, and is difficult to justify theoretically. The method we propose here addresses some of these issues.


\section{Method} \label{method}

The multivariate prediction problem under model \eqref{setup} is equivalent to choosing a decision rule $\delta(x; Y, X) = \{\delta_1(x; Y, X), \ldots, \delta_q(x; Y, X)\}^\top$ to minimize the risk
\begin{equation} \label{eq:CompoundRisk}
  R(B, \delta) = E \left[ \frac{1}{q} \sum_{k=1}^{q} \{ X_0^\top B_k - \delta_k(X_0; Y, X) \}^2 \right],
\end{equation}
where $Y$ and $X$ are $n \times q$ and $n \times p$ matrices containing the training data outcomes and predictors, respectively, and $B_k$ is the $k$th column of the $p \times q$ coefficient matrix $B$.

Our proposed approach is motivated by the fact that minimizing $R(B, \delta)$ \eqref{eq:CompoundRisk} is a compound decision problem, in the sense of \citet{robbins1951asymptotically}. Compound decision problems involve minimizing risk functions that are aggregates of separate individual risk functions; for example, $R(B, \delta)$ is an average of the individual risks from each of the $q$ tasks. A classic compound decision problem is the simultaneous estimation of a vector of normal means under squared error \citep{johnstone2019gaussian}.

A key feature of these problems is that estimators that minimize each of the individual risks over some class of decision rules do not necessarily combine to minimize the aggregate risk \citep{robbins1951asymptotically, stein1956inadmissibility}. A standard example is the James-Stein estimator \citep{james1961estimation}, which dominates the maximum likelihood estimator for the simultaneous estimation of three or more normal means. In the present multivariate prediction context, this phenomenon also explains why the ordinary least squares predictions $X_0^\top (X^\top X)^{-1} X^\top Y$ are suboptimal for minimizing $R(B, \delta)$ \eqref{eq:CompoundRisk}.

One common approach to constructing $\delta$ is to use empirical Bayes methods \citep{efron2019bayes, robbins1985empirical, zhang2003compound}. Following this approach, in \citet{wang2021nonparametric} we assumed the columns of $B$ came from a $p$-dimensional prior, estimated the prior from the training data using nonparametric maximum likelihood, and used the resulting posterior expectations as decision rules. Our results were promising, but our nonparametric estimation scheme suffered for even moderate $p$ and the properties of our procedure were difficult to analyze. Furthermore, a conceptual issue is that the usual connection between compound decision problems and empirical Bayes methods would only hold if the $q$ outcomes were independent, which we do not assume in our model \eqref{setup}.

Here we instead adopt a different strategy to construct $\delta$. We draw from the shrinkage estimation literature, specifically the regression modeling perspective first introduced by \citet{stigler19901988} and recently further studied by \citet{zhao2021regression}. We first interpret miminizing $R(B, \delta)$ \eqref{eq:CompoundRisk} as a regression problem, where the $X_0^\top B_k$ are outcomes, albeit unobserved, and the $\delta_k$ are regression functions. Ideally we would estimate $\delta_k$ nonparametrically, but, as described above, this becomes problematic as $p$ increases. Instead, we consider the class of simple linear models
\begin{equation}
  \label{eq:model}
  \delta_k(X_0; Y, X) = \theta_0 + \sum_{j = 1}^p X_{0j} \hat{B}_{jk} \theta_j,
\end{equation}
where the $\theta_0, \theta_1, \ldots, \theta_p$ are unknown regression parameters, $X_{0j}$ is the $j$th component of $X_0$, and $\hat{B}_{jk}$ is the $j$th component of the ordinary least squares estimate $\hat{B}_k$ of the regression coefficient for the $k$th outcome.

Model \eqref{eq:model} has several interesting implications. First, it can be thought of as using the $X_{0j} \hat{B}_{jk}$ as features to predict the true conditional mean $X_0^\top B_k$. Second, if the $\theta_j$ were all assumed to be equal, model \eqref{eq:model} would be similar to the Efron-Morris estimator for a vector of normal means \citep{efron1973stein}, as it would use a scalar multiple of the $q$-dimensional normal vector $(X_0^\top \hat{B}_1, \ldots, X_0^\top \hat{B}_q)^\top$ to estimate the mean vector $(X_0^\top B_1, \ldots, X_0^\top B_q)^\top$. We allow the $\theta_j$ to differ across the features $j$ to give a more general class of decision rules. Finally, in this regression interpretation, the sample size corresponds to the number of outcomes $q$, so that the more multivariate outcomes we have, the more accurately we can estimate the $\theta_j$. This makes our approach especially well-suited to large-scale multivariate prediction problems.

We aim to estimate the $\theta_j$ that will minimize the compound risk \eqref{eq:CompoundRisk}, which for a decision rule of the form \eqref{eq:model} is
\begin{equation} \label{risk}
  R(B, \Theta) = \frac{1}{q} \sum_{k=1}^{q} E \left\{ \left( 
      X_0^\top B_k - \widetilde{X}_k^\top \Theta
    \right)^2 \right\},
\end{equation}
where $\Theta = (\theta_0, \theta_1, \ldots, \theta_q)^\top$ and
\begin{equation} \label{feature}
  \widetilde{X}_k = \left( 1, X_{01} \hat{B}_{1k} , \ldots , X_{0p} \hat{B}_{pk} \right)^\top.
\end{equation}
While it is natural to try to estimate $\theta_0$ and the $\theta_j$ by minimizing $R(B, \Theta)$ \eqref{risk}, this is impossible because $R(B, \Theta)$ depends on the unknown matrix $B$. On the other hand, we can derive an unbiased estimate of this risk. Let $\hat{\sigma}^2_{0 k}$ be the ordinary least squares estimator of the variance of $X_0^\top \hat{B}_k$ for $k = 1, \ldots, q$. We can show that the empirical risk function
\begin{equation} \label{erisk}
  \hat{R}_q(\Theta) = 
  -\frac{1}{q} \sum_{k=1}^{q} \hat{\sigma}_{0 k}^2 + 
  \frac{1}{q} \sum_{k=1}^{q} \left(X_0^\top \hat{B}_k - \widetilde{X}_k^\top \Theta \right)^2 + 
  2 Q^\top \Theta
\end{equation}
is an unbiased estimate of $R(B, \Theta)$ for each $\Theta$, where 
\begin{equation} \label{Q}
  Q = \left( 0, \frac{1}{q} \sum_{k=1}^{q} X_0^\top \cov(\hat{B}_k) 
    \circ X_0^\top \right)^\top,
\end{equation}
$\cov(\hat{B}_k) = (X^\top X)^{-1}\hat{\sigma}_k$, $\hat{\sigma}_k$ is the ordinary least squares estimate of the conditional variance of the $k$th outcome, and $\circ$ denotes the elementwise product.

\begin{prop}
  \label{prop:unbiased}
  The empirical risk function \eqref{erisk} satisfies $E \hat{R}_q(\Theta) = R(B, \Theta)$.
\end{prop}

We can now propose two estimators that will be studied in the remainder of this paper. We first define the unconstrained estimator
\begin{equation} \label{unconstrained}
  \hat{\Theta}_q 
  =
  \argmin_{\Theta \in \mathbb{R}^{(p+1)}} \hat{R}_q(\Theta) 
  =
  \left( 
    \widetilde{X}^\top \widetilde{X}
  \right)^{-1} \left( \widetilde{X}^\top \hat{B}^\top X_0 - Q \right),
\end{equation}
where $\tilde{X}$ is the $q \times p$ matrix $(\tilde{X}_1, \ldots, \tilde{X}_q)^\top$. This corresponds to the solution of a penalized least squares problem and exists if $p + 1 < q$; the corresponding estimate of $Y_0$ is $\widetilde{X} \hat{\Theta}_q$.

We next define the constrained estimator
\begin{equation} \label{constrained}
  \hat{\Theta}_q^M 
  =
  \argmin_{\Theta \in \mathcal{M}_{q} } \hat{R}_q(\Theta),
  \quad
  \mathcal{M}_q = \{ \Theta : \theta_0 \in [-M_q, M_q], \theta_j \in [0, M_q], j = 1, \ldots, p\},
\end{equation}
where $M_q$ is a constant that can grow with $q$. This estimator constrains the $\theta_1, \ldots, \theta_p$ to be positive. This is sensible because each $\theta_j$ is a scale factor for the ordinary least squares estimate $\hat{B}_{jk}$, so the $\theta_j$ should be positive to ensure that our estimator finds the $j$th feature to be related to the outcome in the same direction as estimated by ordinary least squares. This is similar in principle to the positive-part James-Stein estimator \citep{baranchik1964multiple}. The constant $M_q$ is a technical tool useful for showing uniform convergence in Section \ref{theory}, and in practice can simply be set to a large number. Both methods are implemented in the R package \verb|cole| and available at \url{https://github.com/sdzhao/cole}.

\section{Theoretical properties} \label{theory}

\subsection{Theoretical results for the unconstrained estimator} \label{sec:unconstrained}

We will compare the performance of our unconstrained estimator \eqref{unconstrained} to that of the oracle unconstrained estimator.  Throughout this section, our asymptotics will be in $q$, which is appropriate for large-scale multivariate regression problems. Define the loss function
\begin{equation} \label{eq:loss}
  \ell_q (\Theta) 
  = 
  \frac{1}{q} \sum_{k=1}^{q}\left( X_0^\top B_k - 
    \widetilde{X}_k^\top \Theta \right)^2.
\end{equation}
It is clear that $E \ell_q(\Theta) = R(\Theta)$. Next define the following oracle estimator
\begin{equation} \label{oracle1}
  \hat{\Theta}_q^{\star} 
  =
  \argmin_{\Theta \in \mathbb{R}^{(p+1)}} \ell_q(\Theta) 
  =
  (\widetilde{X}^\top \widetilde{X})^{-1} 
  (\widetilde{X}^\top B^\top X_0),
\end{equation}
which is not feasible because it depends on unknown parameter $B$.

The following theorem shows that our proposed $\hat{\Theta}_q$ \eqref{unconstrained} is close to $\hat{\Theta}_q^{\star}$ \eqref{oracle1}. Define $\vert \Sigma \vert$ to be a $q \times q$ matrix whose entries are equal to the absolute values of the corresponding entries of the conditional covariance matrix $\Sigma$ of the outcomes, from our data-generating model \eqref{setup}.

\begin{theorem}
  \label{theorem1}
  Assume that $X^\top X / n$ and $E(\widetilde{X}^\top \widetilde{X} / q)$ converge to positive definite matrices and let $\lambda_1$ be the largest eigenvalue of $\vert \Sigma \vert$. Then if $\lambda_1 / (q n) \rightarrow 0$,
  \[
    \hat{\Theta}_q - \hat{\Theta}_q^\star = o_P(1).
  \]
\end{theorem}

The assumptions that $X^\top X / n$ and $E(\widetilde{X}^\top \widetilde{X} / q)$ converge are necessary because their dimensions grow with $q$. The quantity $\lambda_1$ is a measure of the strength of the correlations between the different outcomes. In the uncorrelated case where $\Sigma$ is diagonal, the different outcomes provide independent information about $\Theta$ and estimation should be most accurate. This is reflected in the fact that $\lambda_1$ is a constant and $\lambda_1 / (q n) = O\{1 / (qn)\}$ converges quickly to zero. In a perfectly correlated setting where all entries of $\Sigma$ are equal to the same constant, the different outcomes all provide the same information and estimation of $\Theta$ should be most difficult. Indeed, $\lambda_1$ grows with $q$ and $\lambda_1 / (q n) = O(1 / n)$, so that the accuracy of our estimator is driven solely by the sample size of the training data.

The following result shows that the loss of the decision rule using $\hat{\Theta}_q$ is asymptotically as low as that of the rule using $\hat{\Theta}_q^\star$ in probability.

\begin{theorem}
  \label{theorem2}
  If $E Q/q$ converges to a constant and $\lambda_1 / (q n) \rightarrow 0$, then under the conditions of Theorem \ref{theorem1},
  \[
    \ell_q(\hat{\Theta}_q) - \ell_q(\hat{\Theta}_q^\star) = o_P(1).
  \]
\end{theorem}

Because the predicted outcome using the oracle $\hat{\Theta}_q^\star$ will always have lower risk less than or equal to that of the standard ordinary least squares prediction $X_0 (X^\top X)^{-1} X^\top Y$, Theorem \ref{theorem2} shows that our proposed $\hat{\Theta}_q$ will also asymptotically perform no worse.

\subsection{Theoretical results for the constrained estimator} \label{sec:constrained}

Strong results are available for our proposed constrained estimator \ref{constrained}, compared to our unconstrained estimator, because $\Theta$ is restricted to lie in a compact set. First, we can show that the empirical risk function $\hat{R}_q(\Theta)$ \eqref{erisk} is uniformly close to the true loss function $\ell_q(\Theta)$ \eqref{eq:loss} uniformly over $\Theta \in \mathcal{M}_q$.

\begin{theorem}
  \label{theorem3}
  Under assumptions of Theorem \ref{theorem1},
  \[
    \lim_{q \rightarrow \infty}
    E \sup_{\Theta \in \mathcal{M}_q}
    \vert \hat{R}_q(\Theta) - \ell_q(\Theta)\vert = 0,
  \]
  if $\lambda_1 M_q / (q n) \rightarrow 0$, where $\lambda_1$ is the largest eigenvalue of $\vert \Sigma \vert$ defined as in Section \ref{sec:unconstrained}.
\end{theorem}

Theorem \ref{theorem3} requires that the constant $M_q$, which bounds the components of $\Theta \in \mathcal{M}_q$, not grow too quickly, at a rate determined in part by the correlation between the outcomes. For example, if the outcomes are perfectly correlated and $\lambda_1 = O(q)$, $M_q$ must grow no faster than the training data sample size $n$.

Next, define the oracle constrained least squares estimator
\begin{equation} \label{eq:optconstrain}
  \hat{\Theta}_q^{M\star} = \argmin_{\Theta \in \mathcal{M}_{q} } \ell_q(\Theta).
\end{equation}
Then we can show that the expected loss of the proposed constrained estimator \ref{constrained} converges to the expected loss of the oracle constrained estimator \eqref{eq:optconstrain}.

\begin{theorem}
  \label{theorem4}
  If $\lambda_1 M_q / (q n) \rightarrow 0$, then under the conditions of Theorem \ref{theorem3},
  \[
    \lim_{q \rightarrow \infty} \left\{ E \ell_q(\hat{\Theta}_q^M) - E \ell_q(\hat{\Theta}_q^{M\star}) \right\} = 0.
  \]
\end{theorem}

The ordinary least squares prediction of $Y_0$ corresponds to $\theta_0 = 0$ and $\theta_j = 1$ for $j = 1, \ldots, p$, so if $M_q \geq 1$, the predicted outcome using the constrained  $\hat{\Theta}_q^{M \star}$ will always have lower risk less than or equal to that of the standard ordinary least squares prediction, and Theorem \ref{theorem2} shows that our proposed $\hat{\Theta}_q^M$ will also asymptotically perform no worse.

\section{Simulation study} \label{simulation}

\subsection{Settings} \label{settings}
We compared our proposed approaches with four alternative procedures: ordinary least squares, multiresponse group lasso, individual lasso and ridge regressions for each outcome, and the nonparametric empirical Bayes procedure of \citet{wang2021nonparametric}. Ordinary least squares serves as a baseline and corresponds to a naive approach to multivariate regression that does not borrow information across outcomes. We implemented the penalized regression procedures using the R package \verb|glmnet| and tuned using three-fold cross-validation over 50 possible tuning parameters. We implemented the nonparametric empirical Bayes procedure using the R package \verb|cole|, available at \url{https://github.com/sdzhao/cole}. We also considered methods like partial least squares and reduced rank regression, but these performed much worse than the methods we implemented. While other methods for multivariate linear regression problem are also available, for example the tuning-insensitive penalized methods of \citet{liu2015calibrated}, they typically were not implemented in R or were not applicable to large-scale problems where $q \gg n$.

We evaluated the impact of four different factors: correlation between the outcomes, number of outcomes $q$, feature dimension $p$, and sparsity structure of the true parameter matrix $B$. We considered three different structures for $B$: dense, group sparse, and entry sparse. For the dense setting, we generated a $p$-dimensional vector $b \sim N(0, 4 I)$, where $I$ is the $p \times p$ identity matrix, and then let the $k$th column of $B$ equal $b + \tau_k$ for $k = 1, \ldots, q$ and independent $\tau_k \sim N(0, 0.01 I)$. For the group sparse setting, we generated $B$ as in the dense setting and then set every entry to zero except for the first 5 rows. For the entry sparse, we randomly set 60\% of the entries of the dense $B$ to zero.

For each setting, we generated 100 training samples and 50 testing samples following $Y_i = X_i^\top B + \epsilon_i$ \eqref{setup}, where $\epsilon_i \sim N(0, \Sigma)$, with $\Sigma$ was a compound symmetric matrix with diagonal entries equal to 1 and off-diagonal entries equal to a correlation $\rho$ that was varied across replications. We independently generated each component of $X_i$ from a standard normal. We repeated all simulations 100 times and measured performance using the squared error loss \eqref{eq:loss} in the test set, averaged over the replications.

\subsection{Results}
\begin{figure}[h]
  \centering
  \includegraphics[width=0.97\textwidth]{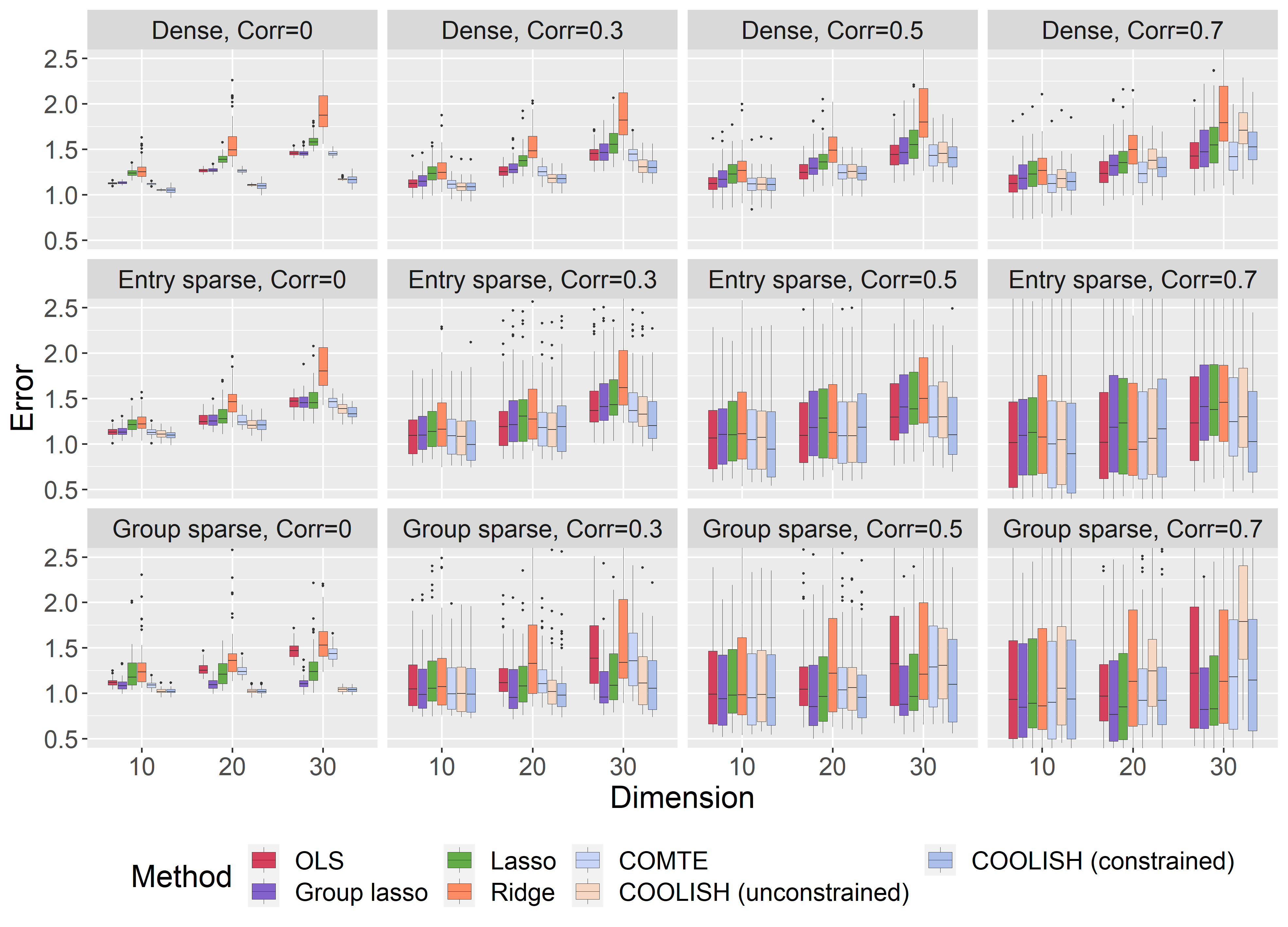}
  \caption{Effect of different outcome correlation levels on average test set errors over 100 replications, with $n = 100$ and $q = 1,000$. \textsc{ols}: ordinary least squares; \textsc{comte}: the method of \citet{wang2021nonparametric}; \textsc{coolish} (unconstrained): the proposed unconstrained estimator \eqref{unconstrained}; \textsc{coolish} (constrained): the proposed constrained estimator \eqref{constrained}.}
  \label{fig1}
\end{figure}

Figure \ref{fig1} shows how the performance of the different methods varied across sparsity structures and correlation values. In this and the following figures we refer to our proposed estimators using the acronym \textsc{coolish}, which stands for COordinate-wise Optimal LInear SHrinkage. These methods typically gave the lowest errors but performed worse as the outcome correlation increased. This is intuitively reasonable given the reliance of our estimators on the effective number of outcomes, as described in Section \ref{theory}. In these settings, our constrained estimator was better than our unconstrained estimator, indicating that the latter may estimate some of the $\theta_j$ to be negative when the correlation is large.

\begin{figure}[h]
  \centering
  \includegraphics[width=0.97\textwidth]{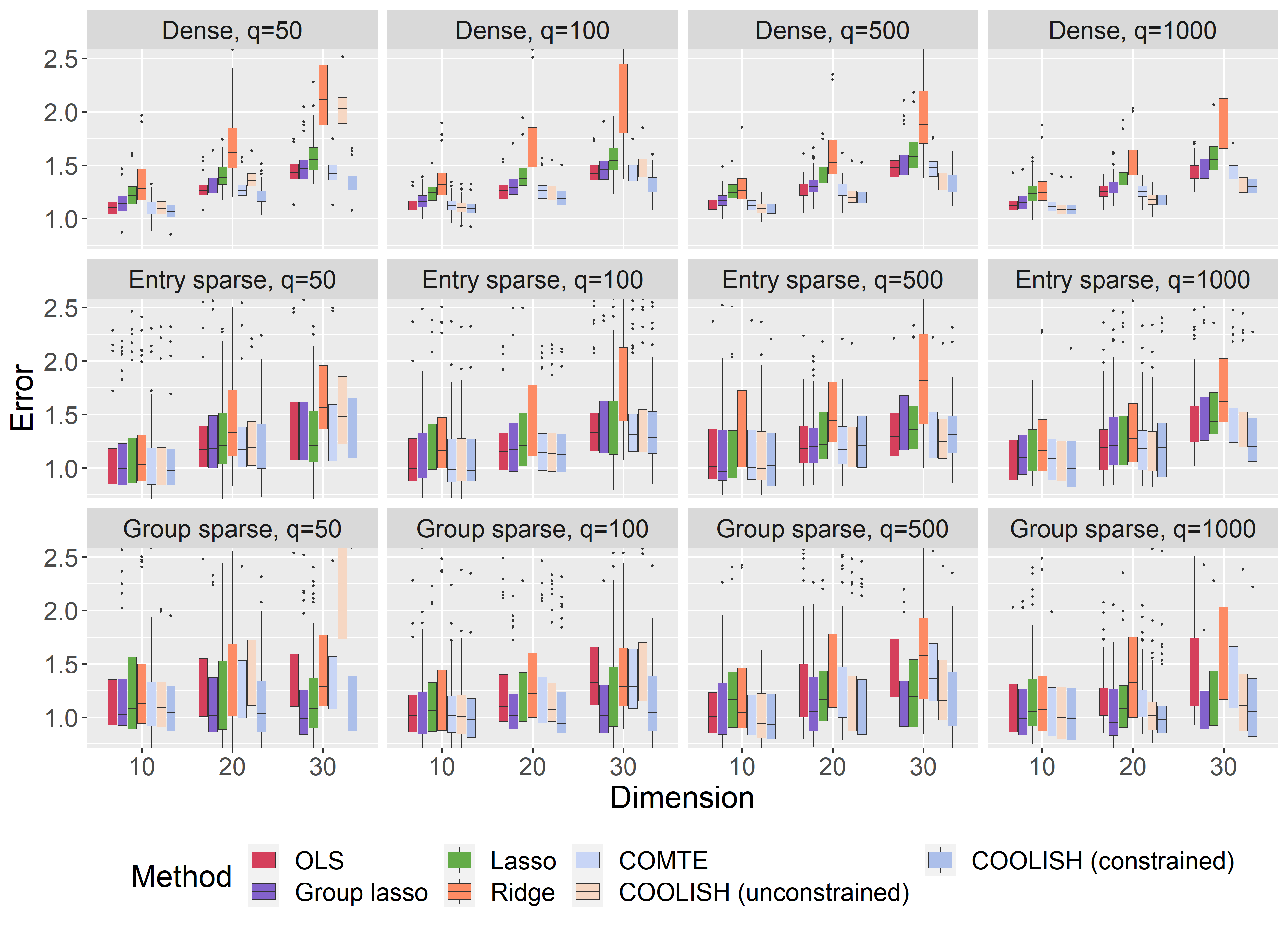}
  \caption{Effect of different $q$ on average test errors over 100 replications, with $n = 100$ and residual error correlations fixed at 0.3. \textsc{ols}: ordinary least squares; \textsc{comte}: the method of \citet{wang2021nonparametric}; \textsc{coolish} (unconstrained): the proposed unconstrained estimator \eqref{unconstrained}; \textsc{coolish} (constrained): the proposed constrained estimator \eqref{constrained}.}
  \label{fig2}
\end{figure}

Figure \ref{fig2} illustrates how performance was affected by different numbers of outcomes $q$, with the correlation between them fixed at 0.3. The proposed unconstrained estimator performed poorly when $q$ was small, which is consistent with the results from Figure \ref{fig1}. In constrast, the proposed constrained method gave the lowest prediction errors among all methods in most cases, and this was apparent even for relatively small $q$.

Both Figures \ref{fig1} and \ref{fig2} show that performance of all methods deteriorated as the feature dimension increased. This was less of an issue for the penalized regression methods, but our proposed methods were affected more heavily because they are built on ordinary least squares estimates, whose accuracy depends heavily on the feature dimension. The figures also show the same performance trends were roughly present in each of the sparsity settings, though our proposed methods showed the greatest relative gain under the dense setting. This was not surprising, as our estimators are linear shrinkage estimators that are best suited for dense parameters.

\section{Data analysis} \label{real}

Single cell RNA-sequencing technologies can measure the expression levels of tens of thousands of genes in individual cells and are revolutionizing genomics research \citep{kolodziejczyk2015technology}. However, a major limitation is that these technologies dissociate cells from each other before sequencing. This step removes information about spatial relationships between cells, which can be key to understanding their function. An emerging alternative technology called MERFISH is able to remedy this issue, and can capture gene expression from individual cells while maintaining their original spatial context \citep{chen2015spatially}. One tradeoff, however, is that MERFISH cannot conveniently interrogate a large number of genes.

A natural question is whether a small number of genes measured using MERFISH can be used to accurately impute the expression levels of the remaining genes \citep{zhu2018identification}. This constitutes a large-scale multivariate regression problem. We study this problem here using two single cell RNA-sequencing datasets derived from honey bee brains \citep{traniello2020meta}, one measuring cells from the whole brain and the other measuring cells from a substructure called the mushroom body. We consider only genes expressed in at least 300 cells in both datasets, which left 3,100 genes measured in 773 cells in the mushroom body dataset and 868 cells in the whole brain dataset. Following \citet{li2018accurate}, we transformed each gene using $\log_{10}(x + 1.01)$, where $x$ was the observed gene expression level after normalizing all cells to have one million reads.

Our goal was to develop a regression model that could use a small set of predictor genes to impute the expression of the rest of the genes. To select the predictor genes, we applied $K$-means clustering to the training data with $K$ equal to the number of features we wanted. In each cluster, we then picked the gene closest to the center of that cluster to serve as a predictor. The number of predictors should equal the number of genes that can be targeted using MERFISH. These numbers are typically chosen to be equal to the number of weight-4 extended Hamming codes that can be constructed using an $n$-digit binary barcode \citep{chen2015spatially}. We therefore studied $K$ = 14, 18, 30, 35, 51, 65, 91, 105, 140, 157, 198, 228, and 285, which correspond to $n = 8, \ldots, 20$.

We first applied the multivariate regression algorithms described in Section \ref{settings} using the mushroom body data as the training set and the whole brain data as the test set, then we reversed the roles of the two datasets. We recorded the average squared prediction errors as well as the computation times of the various methods. We did not implement the method of \citet{wang2021nonparametric} because the residual variances of the ordinary least squares fits were very low and led to computational issues when computing the nonparametric maximum likelihood; this issue is further discussed in \citet{wang2021nonparametric}.

\begin{figure}[h]
  \centering
  \includegraphics[width=0.97\textwidth]{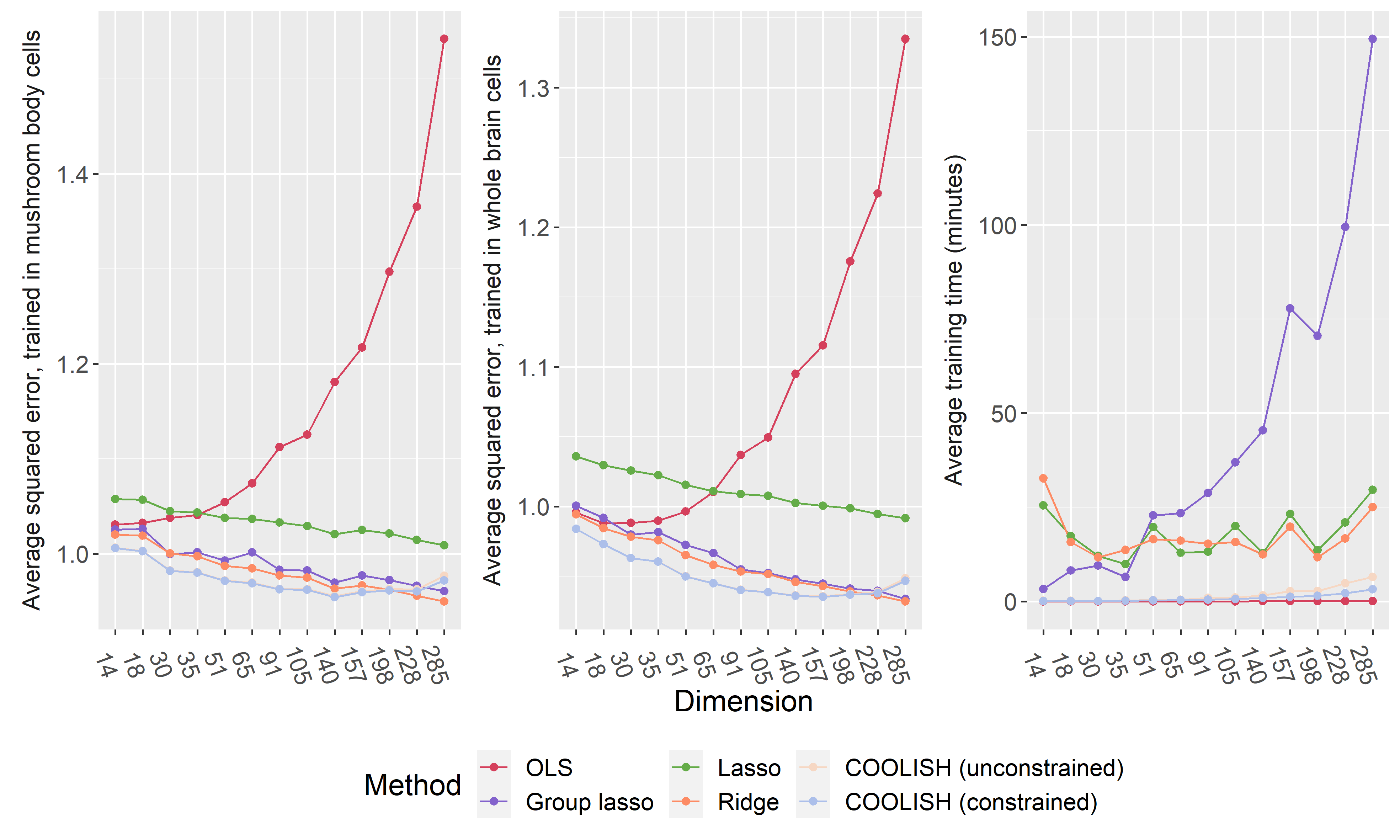}
  \caption{Average prediction errors and computation times in minutes. \textsc{ols}: ordinary least squares; \textsc{comte}: the method of \citet{wang2021nonparametric}; \textsc{coolish} (unconstrained): the proposed unconstrained estimator \eqref{unconstrained}; \textsc{coolish} (constrained): the proposed constrained estimator \eqref{constrained}.}
  \label{figure5}
\end{figure}

The results in Figure \ref{figure5} show that our proposed methods had the best performances for most predictor dimensions. They even outperformed the multiresponse group lasso while requiring just a small fraction of its computation time. Our constrained and unconstrained proposals behaved similarly, with the constrained version performing slightly better for larger $p$. On the other hand, the penalized regression procedures were better when we used more than 228 predictors. This is likely because our estimates are based on ordinary least squares, which suffers as the prediction dimension increases.

\section{Discussion} \label{discussion}

Though our approach was motivated by a regression formulation of shrinkage estimation \citep{stigler19901988, zhao2021regression}, as described in Section \ref{method}, it can also be given an empirical Bayes interpretation. This paradigm would assume that $(X_{01} B_{1k}, \ldots, X_{0p} B_{pk})^\top$ had some prior distribution and then predict the $k$th outcome $Y_{0k}$ using the posterior expectation $E(Y_{0k} \mid X_{01} \hat{B}_{1k} ,\ldots, X_{0p} \hat{B}_{pk})$. If the prior distribution were a $p$-dimensional multivariate normal, the form of the posterior expectation would exactly correspond to our linear model \eqref{eq:model}. In this sense, our approach is similar to the adaptive multivariate ridge regression approach of \citet{brown1980adaptive} except that we target prediction error given a new covariate vector $X_0$, rather than estimation error of the regression coefficients.

A major point of departure between our approach and an empirical Bayes approach is the way in which we estimate linear shrinkage estimator. A typical empirical Bayes approach would first estimate the prior by maximizing the marginal likelihood of the observed vectors $(X_{01} \hat{B}_{1k}, \ldots, X_{0p} \hat{B}_{pk})^\top$ and then use the estimated prior to calculate the posterior expectation. However, these vectors can be arbitrarily dependent across $k$, so the true marginal likelihood cannot be calculated. In contrast, we directly minimize an empirical estimate of the prediction error \eqref{eq:CompoundRisk}, which can also be interpreted as directly estimating the posterior expectation. This simple approach allows us to characterize how the correlation structure of the outcomes affects the asymptotic performance of the decision rule.

We studied only linear models \eqref{eq:model} in this paper, but in principle we could consider more complex regression functions. Allowing the decision rule to be fully nonparametric would be equivalent to the nonparametric empirical Bayes procedure we developed in \citet{wang2021nonparametric}, and would suffer from the curse of dimensionality as the number of predictors increases. Alternatively, we could impose semiparametric models, for which we could use Stein's lemma \citep{stein1981estimation} to obtain empirical risk estimates similar to \eqref{eq:CompoundRisk}. We will study this in future work.

One major limitation of our approach is that we require unbiased and normally distributed estimates of the true $B_{jk}$. These are provided by ordinary least squares estimators in low-dimensional problems, but when the number of features is comparable to or exceeds the sample size, these estimators become less reliable or unavailable. One potential solution may be to use debiased inference techniques \citep{javanmard2014confidence, van2014asymptotically, zhang2014confidence}, but it remains an open problem to develop tuning-free high-dimensional multivariate linear regression algorithms.

\bibliographystyle{abbrvnat}
\bibliography{mybibfile}

\begin{appendix}

\section{Proof of Proposition \ref{prop:unbiased}}
\label{appendix:A}

\begin{align*}
  R(\Theta) 
  &=
    \frac{1}{q} \sum_{k=1}^{q} 
    \mathbb{E} \left(
    X_0^\top B_k - X_0^\top \hat{B}_k 
    + X_0^\top \hat{B}_k - \widetilde{X}_k^\top \Theta
    \right)^2
  \\
  &=
    \frac{1}{q} \sum_{k=1}^{q}
    \mathbb{E} \left\{
    \hat{e}_{0 k}^2 
    + \left(X_0^\top \hat{B}_k - \widetilde{X}_k^\top \Theta \right)^2
    -2 \hat{e}_{0 k} \left(X_0^\top \hat{B}_k - \widetilde{X}_k^\top \Theta \right)
    \right\}
  \\
  &=
    \frac{1}{q} \sum_{k=1}^{q} \sigma_{0k}^2 + 
    \frac{1}{q} \sum_{k=1}^{q} \mathbb{E} 
    \left( X_0^\top \hat{B}_k - \widetilde{X}_k^\top \Theta \right)^2 -
    \frac{2}{q} \sum_{k=1}^{q} \mathbb{E} \hat{e}_{0 k} \left( X_0^\top \hat{B}_k - \widetilde{X}_k^\top \Theta \right)
  \\
  &=
    -\frac{1}{q} \sum_{k=1}^{q} \sigma_{0k}^2 + 
    \frac{1}{q} \sum_{k=1}^{q} \mathbb{E}
    \left( X_0^\top \hat{B}_k - \widetilde{X}_k^\top \Theta \right)^2 +
    2 \mathbb{E} Q^\top \Theta
\end{align*}
The last equation is due to $E\epsilon_k = 0$ and $\cov(\hat{B}_k) = \sigma_{k}^2(X^\top X)^{-1}$. $\hat{\sigma}^2_k = Y^\top_k (I - H)Y_k/(n-p)$ is an unbiased estimator of the variance of $e_k$, $\sigma_k^2$.

\section{Proof of Theorem \ref{theorem1}}
\label{appendix:B}

Let $\hat{Y}_{0 k} = X_0^\top \hat{B}_k$ be the ordinary least squares estimator of $Y_{0 k}$, $\mu_{0 k} = X_0^\top B_{k}$ and $\hat{e}_{0 k} = \hat{Y}_{0 k} - \mu_{0 k}$. Then $\hat{\Theta}_q - \hat{\Theta}_q^\star = ( 
\widetilde{X}^\top \widetilde{X})^{-1}(\widetilde{X}^\top \hat{e}_0 - Q)$, where 

\[
  Q = \left(0, \frac{1}{q} \sum_{k=1}^{q} \sum_{l=1}^{p} X_{0l} \cov(\hat{B}_{1k}, \hat{B}_{lk})X_{01}, \ldots, \frac{1}{q} \sum_{k=1}^{q} \sum_{l=1}^{p} X_{0l} \cov(\hat{B}_{pk}, \hat{B}_{lk})X_{0p} \right)^\top.
\]

If we can show that each component of $1/q (\widetilde{X}^\top \hat{e}_0 - Q)$ converges to zero in probability and each entry of $(\widetilde{X}^\top \widetilde{X}/q)^{-1}$ converges to a constant in probability. Then by Slutsky's theorem, $\hat{\Theta}_q - \hat{\Theta}_q^\star = o_p(1)$.There are two types of terms in $(\widetilde{X}^\top \hat{e}_0 - Q)/q$:
\begin{itemize}
\item $\sum_{k=1}^{q} \hat{e}_{0 k}/q$;
\item $\sum_{k=1}^{q} \hat{e}_{0 k}X_{0 j} \hat{B}_{j k}/q - 
  \sum_{k=1}^{q} \sum_{l=1}^{p} X_{0 l} \cov(\hat{B}_{l k}, \hat{B}_{j k}) X_{0 j}/q$ 
  for $j = 1, \ldots, p$;
\end{itemize}

For the first type of term, the expectation is zero. The variance of the first type of term is
\begin{align*}
  \frac{1}{q^2} \var\left(\sum_{k=1}^{q} \hat{e}_{0 k} \right)
  & = 
    \frac{1}{q^2} \sum_{k, k'}
    \cov\left(\hat{e}_{0 k}, \hat{e}_{0 k'} \right) 
  \\ 
  & =
    \frac{1}{q^2} X_0^\top \left(X^\top X\right)^{-1} X_0 
    \sum_{k, k'} \sigma_{k, k'} 
  \\ 
  & \le 
    \frac{C_{X_0}}{q^2 n} \sum_{k, k'} \left\vert \sigma_{k, k'} \right\vert 
  \\
  & \le
    \frac{\lambda_1 C_{X_0}}{qn} = O(\frac{\lambda_1}{qn}),
\end{align*}
where $\lambda_1$ is the largest eigenvalue of $\vert\Sigma\vert$. $\vert\Sigma\vert$ is taking the element-wise absolute value of $\Sigma$. Let $C_{X_0} = X_0^\top (X^\top X)^{-1} X_0$. For independent case, this rate reduce to $O(1/qn)$ and for perfect dependent case, this reduce to $O(1/n)$.

For the second type of term, the expectation is zero because
\begin{align*}
  &\frac{1}{q} \sum_{k=1}^{q} \hat{e}_{0 k} X_{0 j}\hat{B}_{jk}
    = 
    \frac{1}{q} X_{0j} X_0^\top \left(X^\top X\right)^{-1}
    X^\top E Y^\top X\left(X^\top X\right)^{-1} a_j 
  \\
  \Rightarrow
  &
    \mathbb{E}(\frac{1}{q} \sum_{k=1}^{q} \hat{e}_{0 k} X_{0 j}\hat{B}_{jk})
    =
    \frac{1}{q}  X_{0 j} X_0^\top \left( X^\top X \right)^{-1}a_j 
    \sum_{k=1}^{q} \sigma_k^2,
  \\
  &
    \frac{1}{q} \sum_{k=1}^{q} \sum_{l=1}^{p} X_{0 l} \cov(\hat{B}_{j k}, \hat{B}_{l k}) X_{0j} 
    =
    \frac{1}{q} X_{0 j} X_0^\top \left( X^\top X \right)^{-1}a_j 
    \sum_{k=1}^{q} \hat{\sigma}_k^2
  \\
  \Rightarrow
  &
    \mathbb{E}\left(
    \frac{1}{q} \sum_{k=1}^{q} \hat{e}_{0 k}X_{0 j} \hat{B}_{j k} - 
    \frac{1}{q} \sum_{k=1}^{q} \sum_{l=1}^{p} X_{0 l} \cov(\hat{B}_{j k}, \hat{B}_{l k}) X_{0j}
    \right)
    = 0.
\end{align*} 

By normality assumptions the covariance terms can be simplified as
\begin{align*}
  \cov \left(\hat{e}_{0 k}\hat{B}_{jk}, \hat{e}_{0 k'}\hat{B}_{jk'} \right)
  =&
     B_{jk}B_{jk'}\cov\left( \hat{e}_{0k}, \hat{e}_{0 k'} \right) + 
  \\
   &
     \cov \left(\hat{e}_{0k}, \hat{e}_{0 k'} \right) 
     \cov \left( \hat{B}_{jk}, \hat{B}_{j k'} \right) + 
  \\
   &
     \cov \left(\hat{e}_{0k}, \hat{B}_{j k'} \right) 
     \cov \left(\hat{e}_{0k'}, \hat{B}_{j k} \right) 
  \\
  =& \sigma_{k, k'} B_{jk} B_{jk'} C_{X_0} +
     \sigma_{k, k'}^2 \left( C_{a_j} C_{X_0} + C_{X_0, a_j}^2 \right),
\end{align*}
where $C_{X_0} = X_0^\top (X^\top X)^{-1} X_0$, $C_{a_j} = a_j^\top (X^\top X)^{-1} a_j$ and $C_{X_0, a_j} = a_j^\top (X^\top X)^{-1} X_0$. The variance of the first part of the second type of term is 
\begin{align*}
  \var(\frac{1}{q}\sum_{k=1}^{q} X_{0j} \hat{e}_{0k} \hat{B}_{jk})
  &=
    \frac{X_{0j}^2}{q^2} \sum_{k, k'} \cov\left( \hat{e}_{0k}\hat{B}_{jk}, 
    \hat{e}_{0k'}\hat{B}_{jk'} \right)
  \\
  &=
    \frac{X_{0j}^2}{q^2} \sum_{k, k'} \left\{ 
    \sigma_{k, k'} B_{jk} B_{jk'} C_{X_0} + 
    \sigma_{k, k'}^2  \left(
    C_{a_j}C_{X_0} + C_{X_0, a_j}^2
    \right)
    \right\}
  \\
  &\le
    \frac{X_{0j}^2 C_B^2 C_{X_0}}{q^2} \sum_{k, k'} \left\vert \sigma_{k, k'} \right\vert + 
    \frac{X_{0j}^2 \left(C_{X_0}C_{a_j} + C_{X_0, a_j}^2 \right)}{q^2} \sum_{k, k'} \sigma_{k, k'}^2
  \\
  &= 
    O(\frac{\lambda_1}{qn})
\end{align*}

The last line is an application of Theorem 2 in \cite{guo2019some}. Covariance of the unbiased estimated error variance is 
\begin{align*}
  \cov \left( \hat{\sigma}_k^2, \hat{\sigma}_{k'}^2 \right)
  &=
    \cov \left( \frac{\sum_{i=1}^{n} W_{ki}^2}{n-p}, \frac{\sum_{i=1}^{n} W_{k'i}^2}{n-p} \right)
  \\
  &=
    \frac{1}{(n-p)^2} \sum_{i, i'}^n \left( I-H \right)^2_{i, i'} \sigma_{k, k'}^2,
\end{align*}
where $W_k = (I-H)Y_k$, $I$ is identity matrix and $H = X(X^\top X)^{-1}X^\top$.  Variance of the second part of the second type of terms goes to zero because 
\begin{align*}
  \var\left( \frac{1}{q} \sum_{k=1}^{q} \sum_{l=1}^{p} X_{0 l} \cov(\hat{B}_{j k}, \hat{B}_{l k}) X_{0 j} \right)
  & = \frac{X_{0 j}^2}{(n-p)^2 q^2} \sum_{k, k'}^q \sum_{i, i'}^n
    \left( I - H \right)^2_{i, i'} C_{X_0, a_{j}} \sigma_{k, k'}^2
  \\
  & = 
    O\left( \frac{\lambda_1}{qn^2} \right).
\end{align*}

The covariance of the two parts of the second type of term also converges to zero with the same rate by Cauchy-Schwartz inequality. Thus terms in $(\widetilde{X}^\top \hat{e}_0 - Q)/q$ converges in probability to zero with rate of $O(\lambda_1/qn)$ by Chebyshev's inequality. There are three types of terms in $\widetilde{X}^\top \widetilde{X}/q$:
\begin{itemize}
\item $\sum_{k=1}^q 1/q$;
\item $\sum_{k=1}^q X_{0j} \hat{B}_{jk}/q$;
\item $\sum_{k=1}^q X_{0j} \hat{B}_{jk} X_{0j'} \hat{B}_{j'k}/q$;
\end{itemize}

Because $\mathbb{E}(\widetilde{X}^\top \widetilde{X} / q)$ converges by assumption, the expected value of each type of term converges to a constant. We need to show that variance of each type of term goes to zero. The first type of term is constant with zero variance. For the second type of term
\begin{align*}
  \var\left( \frac{1}{q} \sum_{k=1}^{q} X_{0 j} \hat{B}_{jk} \right) 
  &\le 
    \frac{X_{0j}^2 }{q^2} \sum_{k, k'} \cov\left( \hat{B}_{jk}, \hat{B}_{jk'} \right)
  \\
  &\le
    \frac{X_{0j}^2  C_{a_j}}{q^2} \sum_{k, k'} \left\vert \sigma_{k, k'} \right\vert
  \\
  &=
    O\left( \frac{\lambda_1}{nq} \right).
\end{align*}

For the variance of last type of term, by normality
\begin{align*}
  \cov \left(\hat{B}_{jk} \hat{B}_{j'k}, \hat{B}_{jk'} \hat{B}_{j'k'} \right)
  =&
     B_{jk}B_{jk'}\cov \left(\hat{B}_{j'k}, \hat{B}_{j'k'} \right) +
     B_{jk}B_{j'k'}\cov \left(\hat{B}_{j'k}, \hat{B}_{jk'} \right) + 
  \\
   &
     B_{j'k}B_{jk'}\cov \left(\hat{B}_{jk}, \hat{B}_{j'k'} \right) + 
     B_{j'k}B_{j'k'}\cov \left(\hat{B}_{jk}, \hat{B}_{jk'} \right) + 
  \\
   &
     \cov \left(\hat{B}_{jk}, \hat{B}_{jk'} \right)
     \cov \left(\hat{B}_{j'k}, \hat{B}_{j'k'} \right) +
  \\
   & 
     \cov \left(\hat{B}_{jk}, \hat{B}_{j'k'} \right)
     \cov \left(\hat{B}_{j'k}, \hat{B}_{jk'} \right)
  \\
  =&
     \sigma_{kk'}\left(
     B_{jk}B_{jk'} C_{a_{j'}} + 
     B_{jk}B_{j'k'} C_{a_j, a_{j'}} + \right.
  \\
   &
     \left.
     B_{j'k}B_{jk'} C_{a_j, a_{j'}} + 
     B_{j'k}B_{j'k'} C_{a_j}
     \right) + 
  \\
   &
     \sigma_{k, k'}^2\left(
     C_{a_j}C_{a_{j'}} + 
     C_{a_j, a_{j'}}^2
     \right).
\end{align*}

Thus the variance of the third type of term goes to zero as $\lambda_1/qn$ goes to zero because
\begin{align*}
  \var\left( 
  \frac{1}{q} \sum_{k=1}^q X_{0j} X_{0j'} \hat{B}_{jk} \hat{B}_{j'k} \right)
  =&
     \frac{X_{0j} X_{0j'}}{q^2} \sum_{k, k'} \cov \left(\hat{B}_{jk} \hat{B}_{j'k}, \hat{B}_{jk'} \hat{B}_{j'k'} \right)
  \\
  \le &
	\frac{X_{0j} X_{0j'} C_B^2}{q^2}\left(
	C_{a_{j'}} + C_{a_j} + 2C_{a_j, a_{j'}} \sum_{k, k'} 
	\right)
	\sum_{k, k'} \left\vert \sigma_{k, k'} \right\vert + 
  \\
   &
     \frac{X_{0j} X_{0j'}}{q^2}\left(
     C_{a_j}C_{a_{j'}} + C_{a_j, a_{j'}}^2
     \right)
     \sum_{k, k'}\sigma_{k, k'}^2
  \\
  = &
      O\left(
      \frac{\lambda_1}{qn^2}
      \right).
\end{align*}

Then by Chebyshev's inequality, $\widetilde{X}^\top \widetilde{X}/q - E(\widetilde{X}^\top \widetilde{X}/q) = o_p(\lambda_1/nq)$. Since $E(\widetilde{X}^\top \widetilde{X}/q)$ converges to a positive-definite matrix by assumption, by the continuous mapping theorem $(\widetilde{X}^\top \widetilde{X}/q)^{-1}$ converges to the inverse of $E(\widetilde{X}^\top \widetilde{X}/q)$ in probability.

\section{Proof of Theorem \ref{theorem2}}
\label{appendix:C}

This proof is similar to the proof of Theorem 3.4 in \citet{rigollet2015high}.
The result is clearly true if $\hat{\Theta}_q = \hat{\Theta}_q^\star$. When $\hat{\Theta}_q \neq \hat{\Theta}_q^\star$, since $\hat{R}_q(\hat{\Theta}_q) \le \hat{R}_q(\Theta)$ for any $\Theta$. Let $\hat{Y}_{0 k} = X_0^\top \hat{B}_k$ be the ordinary least squares estimator of $Y_{0 k}$, $\mu_{0 k} = X_0^\top B_{k}$ and $\hat{e}_{0 k} = \hat{Y}_{0 k} - \mu_{0 k}$. Let $\hat{e}_0 = (\hat{e}_{0 1}, \hat{e}_{0 2}, \ldots, \hat{e}_{0 q})^\top$  for $k = 1, 2, \ldots, q$. Then

\[
  \hat{R}_q(\Theta) 
  =
  - \frac{1}{q} \sum_{k = 1}^q \hat{\sigma}_{0 k}^2 + 
  \frac{2}{q} Q^\top \Theta + 
  \frac{1}{q} \sum_{k = 1}^q \hat{e}_{0 k}^2 + 
  \frac{2}{q} \sum_{k = 1}^q \hat{e}_{0 k} (\mu_{0 k} - \widetilde{X}_k^\top \Theta) + \ell_q(\Theta),
\]
it follows that
\[
  0
  \le
  \ell_q(\hat{\Theta}_q) - \ell_q(\hat{\Theta}_q^\star)
  \le
  \frac{2}{q} Q^\top (\hat{\Theta}_q^\star - \hat{\Theta}_q) + 
  \frac{2}{q} \sum_{k = 1}^q \hat{e}_{0 k} \widetilde{X}_k^\top (\hat{\Theta}_q - \hat{\Theta}_q^\star).  
\]

Since $\widetilde{X} (\hat{\Theta}_q - \hat{\Theta}_q^\star) \neq 0$,
\[
  0
  \le
  \ell_q(\hat{\Theta}_q) - \ell_q(\hat{\Theta}_q^\star)
  \le
  \frac{2}{q} Q^\top (\hat{\Theta}_q^\star - \hat{\Theta}_q) + 
  \frac{2}{q} \hat{e}_0^\top \frac{\widetilde{X} (\hat{\Theta}_q - \hat{\Theta}_q^\star)}
  {\Vert\widetilde{X} (\hat{\Theta}_q - \hat{\Theta}_q^\star)\Vert_2} \Vert\widetilde{X} (\hat{\Theta}_q - \hat{\Theta}_q^\star)\Vert_2.
\]
Young's inequality implies that
\[
  \frac{2}{q} \hat{e}_0^\top \frac{\widetilde{X} (\hat{\Theta}_q - \hat{\Theta}_q^\star)}
  {\Vert\widetilde{X} (\hat{\Theta}_q - \hat{\Theta}_q^\star)\Vert_2} \Vert\widetilde{X} (\hat{\Theta}_q - \hat{\Theta}_q^\star)\Vert_2
  \le
  \frac{2}{q} \left\{\hat{e}_0^\top \frac{\widetilde{X} (\hat{\Theta}_q - \hat{\Theta}_q^\star)}
    {\Vert\widetilde{X} (\hat{\Theta}_q - \hat{\Theta}_q^\star)\Vert_2} \right\}^2 + 
  \frac{1}{2 q} \Vert\widetilde{X} (\hat{\Theta}_q - \hat{\Theta}_q^\star)\Vert_2 .
\]

Furthermore,
\begin{align*}
  \ell_q(\hat{\Theta}_q)
  &
    =
    \frac{1}{q} \Vert \mu_0 - \widetilde{X} \hat{\Theta}_q^\star + 
    \widetilde{X} \hat{\Theta}_q^\star - \widetilde{X} \hat{\Theta}_q \Vert_2^2
  \\
  &
    =
    \ell_q(\hat{\Theta}_q^\star) + 
    \frac{2}{q} (\mu_0 - \widetilde{X} \hat{\Theta}_q^\star)^\top
    \widetilde{X} ( \hat{\Theta}_q^\star - \hat{\Theta}_q) + 
    \frac{1}{q} \Vert \widetilde{X} ( \hat{\Theta}_q^\star - \hat{\Theta}_q) \Vert_2^2
  \\
  &
    =
    \ell_q(\hat{\Theta}_q^\star) + 
    \frac{2}{q} \mu_0^\top \left\{ I - \widetilde{X} 
    (\widetilde{X}^\top \widetilde{X})^{-1} \widetilde{X}^\top \right\}
    \widetilde{X} ( \hat{\Theta}_q^\star - \hat{\Theta}_q ) + 
    \frac{1}{q} \Vert \widetilde{X} ( \hat{\Theta}_q^\star - \hat{\Theta}_q) \Vert_2^2
  \\
  &
    =
    \ell_q(\hat{\Theta}_q^\star) + 
    \frac{1}{q} \Vert \widetilde{X} ( \hat{\Theta}_q^\star - \hat{\Theta}_q) \Vert_2^2 .
\end{align*}

Therefore
\begin{equation} \label{t4}
  0 
  \le
  \ell_q(\hat{\Theta}_q) - \ell_q(\hat{\Theta}_q^\star)
  \le
  \frac{4}{q} Q^\top (\hat{\Theta}_q^\star - \hat{\Theta}_q) + 
  \frac{4}{q} \left\{\hat{e}_0^\top \frac{\widetilde{X} 
      (\hat{\Theta}_q - \hat{\Theta}_q^\star)}
    {\Vert\widetilde{X} (\hat{\Theta}_q - \hat{\Theta}_q^\star)\Vert_2} \right\}^2.
\end{equation}

We showed in Theorem \ref{theorem1} that $\var(Q/q) = O(\lambda_1/qn)$. $Q/q - \mathbb{E}(Q/q) $ converges to zero in probability as $\lambda_1/qn$ goes to zero by assumption that $\mathbb{E}(Q/q)$ converges to a constant and Chebyshev's inequality. Since $\hat{\Theta}_q - \hat{\Theta}_q^\star $ converges to zero in probability as $\lambda_1/qn$ goes to zero by Theorem \ref{theorem1}, the first term in \eqref{t4} converges to zero in probability as $\lambda_1/qn$ goes to zero. 

To show that the second term in \eqref{t4} is $o_p(1)$, let $\Phi$ be a $q$ by $(p+1)$ matrix whose columns constitute an orthonormal basis of the column space of $\widetilde{X}$, as in the proof of Theorem 2.2 of \citet{rigollet2015high}. Then there exists a $\nu \in \mathbb{R}^{p+1}$ such that $\Phi \nu = \widetilde{X} (\hat{\Theta}_q - \hat{\Theta}_q^\star)$. Therefore

\[
  \frac{1}{q} \left\{\hat{e}_0^\top \frac{\widetilde{X} 
      (\hat{\Theta}_q - \hat{\Theta}_q^\star)}
    {\Vert\widetilde{X} (\hat{\Theta}_q - \hat{\Theta}_q^\star)\Vert_2} \right\}^2
  =
  \frac{1}{q} \left( \hat{e}_0^\top \frac{\Phi \nu}{\Vert\Phi \nu\Vert_2} \right)^2
  =
  \frac{1}{q} \left( \hat{e}_0^\top \Phi \frac{\nu}{\Vert\Phi \nu\Vert_2} \right)^2
  \le 
  \left(\frac{1}{q^{1/2}} \sup_{ u \in \mathcal{B}_1} 
    \vert \hat{e}_0^\top \Phi  u \vert \right)^2 ,
\]
where $\mathcal{B}_r$ is the closed ball in $\mathbb{R}^{p+1}$ of radius $r$ about the origin. Using the arguments in the proof of Theorem 1.19 of \citet{rigollet2015high}, 
\[
  P \left( \sup_{ u \in \mathcal{B}_1} 
    \vert \hat{e}_0^\top \Phi  u \vert \ge q^{1/2} t^{1/2} \right)
  \le
  P \left( 2 \sup_{ u \in \mathcal{N}} 
    \vert \hat{e}_0^\top \Phi  u \vert \ge q^{1/2} t^{1/2} \right),
\]
for any $t > 0$, where $\mathcal{N}$ is an $1/2$-net of $\mathcal{B}_1$. by Markov's inequality
\[
  P \left(\vert \hat{e}_0^\top \Phi u \vert \ge q^{1/2} t^{1/2} / 2 \right)
  \le 
  \frac{4 \mathbb{E} \{ (\hat{e}_0^\top \Phi u )^2 \}}
  {q t} .
\]

Let $c_j$ be the $j$th coordinate of $\Phi u$ and $c = (c_1, \ldots, c_q)^\top$. $c^\top c = 1$ because $\Vert \Phi u \Vert = 1$. Then 
\begin{align*}
  \frac{1}{q} \mathbb{E} \{ (\hat{e}_0^\top \Phi u)^2 \}
  &=
    \frac{1}{q} \mathbb{E} \left\{ \sum_{ij}^q \hat{e}_{0 i} c_i \hat{e}_{0 j} c_j 
    \right\}
  \\
  &= 
    \frac{C_{X_0}}{q} c^\top \Sigma c
  \\
  &\le
    \frac{C_{X_0} \lambda_1(c^\top c)}{q}
    = 
    O\left(
    \frac{\lambda_1}{qn}
    \right).
\end{align*}

Since the $1/2$-net $\mathcal{N}$ has cardinality at most $6^{p+1}$ by Lemma 1.18 of \citet{rigollet2015high},
\[
  P \left( \sup_{ u \in \mathcal{B}_1} 
    \vert \hat{e}_0^\top \Phi  u \vert \ge q^{1/2} t^{1/2} \right)
  \le 
  \frac{6^{p+1} 4 \mathbb{E} \{ (\hat{e}_0^\top \Phi u )^2 \}}
  {q t} 
  \rightarrow
  0,
\]
as $\lambda_1/qn$ goes to zero. Therefore
\[
  P \left[ \frac{1}{q}\left\{ \hat{e}_0^\top \frac
      {\widetilde{X}(\hat{\Theta}_q - \hat{\Theta}_q^\star)}
      {\Vert\widetilde{X} (\hat{\Theta}_q - \hat{\Theta}_q^\star)\Vert_2}
    \right\}^2 > t \right]
  \rightarrow
  0 .
\]

For every $t > 0$, which implies that $\ell_q(\hat{\Theta}_q) - \ell_q(\hat{\Theta}_q^\star) \xrightarrow[]{P} 0$ as $\lambda_1/qn$ goes to zero. 

\section{Proof of Theorem \ref{theorem3}}
\label{appendix:D}

Let $\hat{Y}_{0 k} = X_0^{\top} \hat{B}_k$ and $\mu_{0 k} = X_0^{\top} B_k$. Then $\hat{Y}_{0 k}$ follows normal distribution with mean $\mu_{0 k}$ and variance $\sigma_{0 k}^2$. Let $\hat{\sigma}_{0 k}^2$ be the ordinary least squares estimator of $\sigma_{0 k}^2$.
Let $\hat{e}_{0 k} = \hat{Y}_{0 k} - \mu_{0 k}$. Since 
\[
  \frac{1}{q} \sum_{k = 1}^{q} (\hat{Y}_{0 k} - 
  \Theta^\top\widetilde{X}_k)^2 
  = 
  \frac{1}{q} \sum_{k = 1}^q \left\{ \hat{e}_{0 k}^2 + 2 \hat{e}_{0 k}(\mu_{0 k} - 
    \Theta^\top \widetilde{X}_k) + (\mu_{0 k} - 
    \Theta^\top \widetilde{X}_k )^2 \right\}.
\]

It follows that
\begin{align} \label{eq:t1p1}
  \nonumber
  & 
    \underset{\Theta \in \mathcal{M}_q}{\text{sup}} \vert\hat{R}_q(\Theta) - 
    \ell_q(\Theta)\vert 
    = 
    \underset{\Theta \in \mathcal{M}_q}{\text{sup}} 
    \left\vert-\frac{1}{q}\sum_{k = 1}^q \hat{\sigma}_{0 k}^2 + 
    \frac{2}{q} Q^\top\Theta + 
    \frac{1}{q} \sum_{k = 1}^q (\hat{e}_{0 k}^2 + 
    2\hat{e}_{0 k} \mu_{0_k} - 
    2\hat{e}_{0 k} \widetilde{X_k}^\top \Theta ) \right\vert  
  \\
  \le 
  &
    \left\vert\frac{1}{q} \sum_{k = 1}^q \hat{\sigma}_{0 k}^2 - 
    \frac{1}{q} \sum_{k = 1}^q \hat{e}_{0 k}^2 \right\vert + 
    \left\vert\frac{2}{q} \sum_{k = 1}^q \hat{e}_{0 k} \mu_{0 k}
    \right\vert + 2\underset{\Theta \in \mathcal{M}_q}{\text{sup}}
    \left\vert \frac{1}{q} (Q - \sum_{k = 1}^q \hat{e}_{0 k} 
    \widetilde{X}_k)^\top \Theta \right\vert,
\end{align}
where $Q$ is defined in \eqref{Q}. The next step is to show terms in \eqref{eq:t1p1} converge to zero in expectation. For the first term in \eqref{eq:t1p1},
\begin{align} \label{eq: cov1}
  \left( E \left\vert \frac{1}{q} \sum_{k = 1}^q 
  \hat{\sigma}_{0 k}^2 - \frac{1}{q} \sum_{k = k}^q
  e_{0 k}^2 \right\vert \right)^2 
  \le&
       E \left[ \left\{ \frac{1}{q} \sum_{k = 1}^q (\hat{\sigma}_{0 k}^2 - e_{0 k}^2) \right\}^2 \right] \nonumber 
  \\
  =&   
     \frac{1}{q^2} \sum_{k, k'}^q \operatorname{cov}(\hat{\sigma}_{0 k}^2 - 
     e_{0 k}^2, \hat{\sigma}_{0 k'}^2 - e_{0 k'}^2) 
     \nonumber
  \\
  \le&
       \frac{1}{q^2} \sum_{k \neq l}^q 
       \vert\cov ( \hat{ \sigma}_{0 k}^2, \hat{\sigma}_{0 l}^2)\vert 
       + 
       \frac{1}{q^2} \sum_{k \neq l}^q 
       \vert\cov ( e_{0 k}^2, e_{0 l}^2)\vert 
       \nonumber 
  \\
     &+ 
       \frac{2}{q^2} \sum_{k \neq l}^q 
       \vert\cov( \hat{ \sigma}_{0 k}^2, e_{0 l}^2)\vert.
\end{align}

Let $C_{X_0} = X_0^{\top} (X^\top X)^{-1} X_0$, $H = X(X^\top X)^{-1}X^\top$ and $Z_k = (I - H) Y_k$. Then $\hat{\sigma}_{0 k}^2 = C_{X_0} \frac{\sum_{i = 1}^n Z_{k i}^2 }{n-p}$. Let $C_H$ be the largest element in $(I - H)$. The first part in \eqref{eq: cov1} is bounded by
\begin{align*}
  \frac{1}{q^2} \sum_{k, k'}^q 
  \vert\cov (C_{X_0} \frac{\sum_{i=1}^n Z_{k i}^2 }{n-p}, 
  C_{X_0} \frac{\sum_{i=1}^n Z_{k' i}^2 }{n-p}) \vert
  &\leq  \frac{C_{X_0}^2}{(n-p)^2 q^2} \sum_{k, k'}^q 
    \sum_{i, i'}^n \vert\cov( Z_{k i}^2, Z_{k' i'}^2)\vert 
  \\
  &
    = 
    \frac{2 C_{X_0}^2}{(n-p)^2 q^2} \sum_{k, k'}^q 
    \sum_{i, i'}^n (I - H)_{i i'}^2 \sigma_{k, k'}^2
  \\
  &
    \leq
    \frac{2 C_{X_0}^2 C_H^2}{q_2} 
    \sum_{k, k'}^q \sigma_{k, k'}^2 
    = O\left(\frac{\lambda_1}{qn^2}\right).
\end{align*}

For two mean zero normal random variable $\hat{e}_{0 k}$ and $\hat{e}_{0 k'}$, $\cov(\hat{e}_{0 k}, \hat{e}_{0 k'}) = 2 \cov(\hat{e}_{0 k}, \hat{e}_{0 k'})^2$. Thus the second term in \eqref{eq: cov1} is bounded by $\frac{2 C_{X_0}^2}{q^2} \sum_{k k'}^q \sigma_{k k'}^2 = O(\lambda_1/qn^2)$. Let $K_H = \max_{i} (X_0^\star (X^\top X)^{-1} X^\top(I-H)_i)^2 $ where $(I-H)_i$ is the $i$th column of $(I-H)$ for $i = 1, 2,\ldots, n$. The last part in \eqref{eq: cov1} is at most
\begin{align*}
  &
    \frac{2}{q^2} \sum_{k, k'}^q 
    \vert\cov ( C_{X_0} \frac{\sum_{i = 1}^n Z_{k i}^2 }{n-p}, e_{0 k'}^2)\vert
    =
    \frac{2C_{X_0}}{(n-p) q^2} \sum_{k, k'}^q  
    \sum_{i = 1}^n \vert\cov (Z_{k i}^2, e_{0 k'}^2)\vert
  \\
  = 
  &
    \frac{4C_{X_0}}{(n-p) q^2} \sum_{k, k'}^q  \sum_{i = 1}^n
    \left( X_0 (X^\top X)^{-1} X^\top(I-H)_i \sigma_{k, k'} \right)^2
  \\
  \le
  &
    \frac{4C_{X_0} n K_H}{(n-p) q^2} \sum_{k, k'}^q \sigma_{k, k'}^2 = O\left(
    \frac{\lambda_1}{qn}
    \right).
\end{align*}

The second term in \eqref{eq:t1p1} obeys
\begin{align*}
  \left( E\left\vert \frac{2}{q} \sum_{k = 1}^q 
  \hat{e}_{0 k} \mu_{0 k} \right\vert \right)^2 
  &\le 
    \frac{4}{q^2} 
    E\left\{ \left( \sum_{k = 1}^q \hat{e}_{0 k} 
    \mu_{0 k} \right)^2  \right\}
  \\
  &
    =
    \frac{4}{q^2}  \sum_{k, k'}^q 
    \mu_{0 k} \mu_{0 k'} \cov (\hat{e}_{0 k} , \hat{e}_{0 k'})
  \\
  &
    \le
    \frac{4C_{X_0}^2 K_{\mu}^2}{q^2} \sum_{k, k'}^q 
    \left\vert \sigma_{k, k'} \right\vert = O\left(\frac{\lambda_1}{qn^2}\right). 
\end{align*}

For the last part in \eqref{eq:t1p1}, let $M = \sup_{\Theta \in \mathcal{M}_q} \Vert\Theta\Vert_{\infty}$.
\begin{align*}
  \left\{\mathbb{E}\underset{\Theta \in \mathcal{M}_q}{\text{sup}}
  \left\vert \frac{1}{q} (Q - \sum_{k = 1}^q \hat{e}_{0 k} 
  \widetilde{X}_k)^\top \Theta \right\vert
  \right\}^2 
  &
    \le
    M^2 \mathbb{E} \left\{
    \sum_{l=1}^{p+1} \frac{1}{q} \left(
    Q - \widetilde{X}^\top \hat{e}_0
    \right)_l
    \right\}^2
  \\
  &
    =
    M^2 \frac{1}{q^2} \sum_{l, l'}^{p+1}
    \cov\left\{
    \left(
    Q - \widetilde{X}^\top \hat{e}_0 
    \right)_l,
    \left(
    Q - \widetilde{X}^\top \hat{e}_0
    \right)_{l'}
    \right\} \\
  &
    =
    O\left(
    \frac{\lambda_1 M^2}{qn}
    \right)
\end{align*}

The first equation above is because the expectations of terms in $Q - \widetilde{X}^\top \hat{e}_0$ is zero as showed in Theorem \ref{theorem1}. We also showed in Theorem \ref{theorem1} that variances of elements in $Q - \widetilde{X}^\top \hat{e}_0 = O(\lambda/qn)$. The last line follows by applying Cauchy-Schwartz inequality.

\section{Proof of Theorem \ref{theorem4}}
\label{appendix:E}

Let $\hat{\Theta}_q^M$ be defined as \eqref{constrained}, $\hat{\Theta}_q^{M\star}$ be defined as \eqref{eq:optconstrain} and $\ell_q$ be defined as \eqref{eq:loss}. Because both $\hat{\Theta}_q^M$ and $\hat{\Theta}_q^{M\star}$ lie in $\mathcal{M}_q$
\begin{align*}
  0 
  \le
  \ell_q(\hat{\Theta}_q^M) - \ell_q(\hat{\Theta}_q^{M\star}) 
  =
  &
    \ell_q(\hat{\Theta}_q^M)  - \hat{R}_q(\hat{\Theta}_q^M) + 
    \hat{R}_q(\hat{\Theta}_q^M) - \hat{R}_q(\hat{\Theta}_q^{M\star}) + 
    \hat{R}_q(\hat{\Theta}_q^{M\star}) - \ell_q(\hat{\Theta}_q^{M\star}) 
  \\
  \le
  &
    2 \underset{\Theta \in \mathcal{M}_q}{\text{sup}} \vert \ell_q (\Theta)  -
    \hat{R}_q( \Theta ) \vert + \hat{R}_q( \hat{\Theta}_q^M ) - 
    \hat{R}_q( \hat{\Theta}_q^{M\star} ) . 
\end{align*}

By construction, $\hat{R}_q(\hat{\Theta}_q^M) \le \hat{R}_q(\hat{\Theta}_q^{M\star})$, so
\[
  0 
  \le 
  E\left\{ \ell_q(\hat{\Theta}_q^M) - \ell_q(\hat{\Theta}_q^{M\star})  \right\} 
  \le
  2E \underset{\Theta \in \mathcal{M}_q}{\text{sup}} 
  \vert \ell_q(\Theta)  - \hat{R}_q(\Theta) \vert,
\]
which converges to zero by Theorem \ref{theorem3} if $\lambda_1 M/qn$ converges to zero.
\end{appendix}

\end{document}